\documentstyle[11pt]{article} 
\def\bas\def\baselinestretch{1.2} 
\def\ltsim{\lower3pt\hbox{$\, \buildrel < \over \sim \, $}} 
\def\gtsim{\lower3pt\hbox{$\, \buildrel > \over \sim \, $}} 
\def\indic{\lower3pt\hbox{$\, \buildrel (m_1,m_2)\not=0 \over 
\kappa_{\vec m}\ll 1 \, $}} 
\input{psfig} 

\begin{document} 
  
\begin{flushright}  
{OUTP-2000-01-P}\\   
\end{flushright}  
\vskip 0.5 cm 
 
\begin{center} 
{\Large {\bf On gauge unification in Type I/I$^{\prime}$ models \bigskip }} 
\\[0pt] 
\bigskip {\large {\bf Dumitru M. Ghilencea\footnote{{{ {\ {\ {\ 
E-mail: D.Ghilencea1@physics.oxford.ac.uk}}}}}} } and {\bf Graham G.
Ross\footnote{{{ {\ {\ {\ E-mail: G.Ross1@physics.oxford.ac.uk}}}}}} 
\bigskip }}\\[0pt] 
{\it Department of Physics, Theoretical Physics, University of 
Oxford}\\[0pt] 
{\it 1 Keble Road, Oxford OX1 3NP, United Kingdom}\\[0pt] 
\bigskip 
 
\vspace{3cm} $Abstract$ 
\end{center} 
 
{\small We discuss whether the (MSSM) unification of gauge couplings can be 
accommodated in string theories with a low (TeV) string scale. This requires 
either power law running of the couplings or logarithmic running 
extremely far above the string scale. In both cases it is 
difficult to arrange for the multiplet structure to give the MSSM 
result. For the case of power law running there is also enhanced sensitivity 
to the spectrum at the unification scale. For the case of logarithmic 
running there is a fine tuning problem associated with the light closed 
string Kaluza Klein spectrum which requires gauge mediated supersymmetry 
breaking on the ``visible'' brane with a dangerously low scale of 
supersymmetry breaking. Evading these problems in low string scale models
requires a departure from the MSSM structure, which would
imply that the success of gauge unification in the MSSM is just an 
accident.} 
\newpage  
\setcounter{page}{1}
\section{Introduction} 
 
Perhaps the most persuasive indication of structure ``Beyond the Standard 
Model'' comes from the observation that the strong, and electroweak gauge 
couplings unify in the simplest supersymmetric extension of the Standard 
Model. Computing the radiative corrections to the gauge couplings coming 
from the virtual states of the MSSM one finds the $SU(3),SU(2)$ and $U(1)$ 
couplings become very nearly equal at the scale $(1-3). 10^{16}$ GeV 
provided 
one adopts the $SU(5)$ normalisation of the $U(1)$ factor. A measure of the 
accuracy of this prediction may be obtained by using the assumed unification 
to predict the value of the strong coupling given the weak and 
electromagnetic couplings.
 One finds $\alpha_{3}(M_z)=0.126\pm 0.003$ to be 
compared with the experimental value \cite{particledata} 
$\alpha_{3}(M_z)=0.119\pm 0.002$. Only 
slightly less impressive is the fact that the SUSY radiative
corrections, assuming the same unification scale, lead to a 
prediction of the bottom mass in the range 
$3.12<m_b(M_z)<3.76 GeV$ 
to be compared to the range $2.72< m_b(M_z)< 3.16 GeV$ obtained from
the pole mass $M_b=4.8\pm 0.2 GeV$ \cite{pokorski}.
Again this prediction assumes the $SU(5)$ prediction for the 
equality of the bottom to the tau mass at the unification scale. It is 
largely on the basis of these successes and the fact that supersymmetry is 
needed to solve the mass hierarchy problem that there has been so much 
interest in supersymmetric models and in compactification schemes that 
preserve a low energy supersymmetry. 
 
Weakly coupled heterotic string compactifications accommodate these 
features in a very natural way. The string and gauge unification
scales are predicted 
to be very large, of the order of the Planck scale. The $SU(5)$ 
normalisation of the $U(1)$ factor emerges for a wide variety of 
compactifications even without a stage of Grand Unification below the Planck 
scale. The relation of the bottom to tau masses is more model dependent but, 
because there is an underlying $E_{6}$ symmetry above the compactification 
scale, it is easy to find specific models which do predict equality even 
without a stage of Grand Unification. 
 
This picture has been challenged by the realisation that it is possible to 
build (open) string theories in which the string scale is much smaller
\cite{savas}-\cite{lykken}, possibly no more than the TeV scale which 
might be 
small enough to avoid the gauge hierarchy problem. Associated with this is 
the possibility that there are new large dimensions, possibly as large as a 
millimetre. At first sight such theories seem to be inconsistent with the 
MSSM unification of couplings which requires a very large unification scale. 
However there are several possible ways that gauge unification may be 
maintained. One is that some, or all, of the Standard Model states may 
propagate in new large dimensions, leading to power law running of the gauge 
couplings and the possibility that unification is achieved at a much lower 
scale. A second possibility \cite{bachas} is that the normal logarithmic 
running applies but the ultraviolet cut-off scale is not the string scale 
but is associated with a very heavy Kaluza Klein or winding mode. This is a 
particularly interesting possibility because the cut-off scale is related to 
the {\it infra-red} properties of the transverse (closed string) channel. 
The mass, $m_{KK},$ of the lightest Kaluza Klein states in this channel is 
inversely proportional to the new large compactification radius and is
of 
$O(10^{-4}eV)$ for a compactification radius of $O(1mm).$ A state of this 
mass corresponds to a cutoff scale in the open string channel of 
$O(M_{s}^{2}/m_{KK})$ and if one sets the string scale, $M_{s},$ to 1 TeV one 
obtains $10^{17}GeV$ for the ultra violet cut-off. This is tantalisingly 
close to the gauge unification scale in the MSSM. 
 
In this letter we will consider both these possibilities for the case of 
Type I/I$^{\prime }$ string compactifications. We will concentrate on the 
question whether it is possible to maintain the remarkable success of the 
MSSM predictions in the sense that the running of the gauge couplings is 
governed by the MSSM beta functions. Our conclusion is somewhat pessimistic 
suggesting that MSSM unification with a low string scale does not follow as 
nicely as it did in the heterotic string case. Of course it may be that 
nature is perverse and that the success of unification in the MSSM could
be just an accident. As an illustration of this we also consider a 
promising Type I  string compactification in which 
gauge unification can be made to work. 
 
\section{Type I/I$^\prime$ models and unification with low string scale} 
 
In Type I (Type I$^{\prime }$) strings at weak coupling the Standard Model 
fields are described by open strings confined to a $D_{p}$ brane with Kaluza 
Klein modes with respect to $(p-3)$ compact dimensions in the brane and 
winding modes with respect to $(9-p)$ dimensions orthogonal to the brane. 
Closed strings mediate the gravitational interaction and are free to 
propagate in the full ten dimensional space-time. Upon compactification to 
four dimensional space-time, the following relations emerge between the 
volume\footnote{in Type I/I$^{\prime }$ string length 
units.} $v_{p}$ of the compact 
dimensions of the SM brane, the volume $v_{o}$ of the dimensions orthogonal 
to it and the string scale $M_{I}$  
\begin{equation} 
M_{I}\sim g^{2}M_{P}\left( {\frac{v_{p}}{v_{o}}}\right) ^{\frac{1}{2}} 
\label{M_I} 
\end{equation} 
so the string scale may be reduced \cite{witten} and may be very low (``TeV 
scale'') \cite{lykken} if the transverse space is large \cite{savas}. It 
should however be noted that lowering the fundamental scale does not, by 
itself, solve the hierarchy problem for now we must explain why $v_{o}\gg 
v_{p}$. 
In addition to the relation of eq.(\ref{M_I}) the string coupling is related 
to the four dimensional gauge coupling  
\begin{equation} 
\lambda _{I}\sim g^{2}v_{p}  \label{weak} 
\end{equation} 
where $v_{p}$ should be of order unity to keep $\lambda _{I}<1$. 
 
Particularly interesting is the case in which the MSSM is embedded in a $D3$ 
brane, for in this situation one may have extra (transverse) dimensions as 
large as $\leq mm$ \cite{savas} (not ruled out by experiment) while the 
Standard Model field has only winding modes corresponding to the large 
transverse dimensions. In the case the MSSM is embedded in a $D_{p}$ brane 
with $p>3$ the $p-3$ additional dimensions cannot be much larger than 
$1\,TeV^{-1}$ due to the presence of Standard Model Kaluza 
Klein modes. However 
in asymmetric compactifications some or all of the $(9-p)$ orthogonal 
dimensions may be much larger. 
 
Unlike Grand Unification, string theories predict the gauge unification 
scale. In the heterotic string the Standard Model fields are closed string 
states and the ultraviolet cutoff scale is determined by the geometry of the 
string world sheet. At one loop this is the torus and one may readily verify 
that the cutoff is at the string scale \cite{dixon,string_thresholds}. In the 
case the Standard Model fields are described by open strings however the 
world sheet geometry does not regulate the divergences the latter only being 
eliminated after summing over the possible geometries. After eliminating the 
divergences the UV behaviour in the open string channel is determined by the  
IR behaviour in the transverse open string channel. This follows due 
to the closed string/open string duality and leads to alternative, but 
equivalent, descriptions of the behaviour of the radiative corrections to 
gauge couplings. The alternative open and closed string descriptions are 
summarised in  Table~1. 
 
Looking at the first two entries in Table 1 one may see that in the open 
string channel the running of the couplings is due to the familiar radiative 
corrections to the gauge couplings while in the closed string channel it is 
due to the modification of the vacuum expectation value of the bulk
field, 
$\phi ,$ which is the coefficient of the gauge kinetic term. The latter 
running is due to the $\phi $ propagator in the bulk, emitted by distant 
branes, evaluated at the transverse position of the SM brane. This provides 
a classical interpretation of the quantum effects on the gauge couplings (UV 
behaviour) on the SM brane. Logarithmic evolution in the direct channel is 
due to the radiative corrections involving the propagation of virtual 
Standard Model states in four dimensions. In the transverse channel this is 
due to $\phi $ propagation in two transverse dimensions. If the open string 
states propagate in more than four (``in-brane'') dimensions 
the couplings will run like a 
power rather than a logarithm. 
 
From the four dimensional point of view an (``in-brane'')
extra dimension shows up as a 
Kaluza Klein (KK) tower of excitations which each contribute additional 
logarithmic running to the coupling which sum to a (linear) 
power once sufficient KK 
states can be excited. In the transverse closed string channel this 
corresponds to $\phi $ propagation in one transverse dimension (the Green 
functions fall as $r^{2-d_{\perp }}$ for $d_{\perp }$ transverse 
dimensions). Similarly if there are two KK towers of excitations in the 
direct channel the couplings evolve quadratically. In the closed string 
channel this corresponds to the case $d_{\perp }=0.$ In the weakly coupled 
heterotic string with orbifold or Calabi Yau supersymmetric 
compactifications from 10 to 4 dimensions it might seem there could be 
running proportional to $r^{6}.$ However, the massive KK modes giving rise to 
the extra dimensions fill out $N=2$ and $N=4$ supermultiplets. The latter 
do not contribute to radiative corrections and so the power law running is 
limited to $r^{2}.$ In the transverse channel this result follows simply 
from the observation that $d_{\perp }=0$ is the limiting case. 
 
The prediction of gauge unification also requires that the tree level 
coupling constant ratios be predicted. The gauge couplings are related to 
the dilaton couplings to the various gauge kinetic terms. In the heterotic 
string the reason for unification is the presence of a single dilaton v.e.v. 
which is the string coupling at the unification scale. For level-1 Kac-Moody 
theories one has the usual $SU(5)$ relations. More generally such a relation 
may come from an underlying Grand Unified symmetry above the 
compactification scale. In Type I/I$^\prime$ theories 
it has been suggested that in the closed string channel 
\cite{jmr}  it is an underlying geometric symmetry of the brane sector 
that is responsible for relating the couplings. 
 
Finally string theories predict the gauge unification scale. The prediction 
for the tree level ratios apply at the scale at which the radiative 
corrections are cut-off. As discussed above this is determined by the 
lightest relevant state in the transverse channel. This turns out either to 
be the string scale or the lightest KK scale depending on which sector of 
the theory one is considering \cite{dudash}. 
 
\begin{table}[tbp] \centering%
%
\begin{tabular}{|p{2in}|p{2in}|p{2in}|} 
\hline 
& {\bf Open string channel} & {\bf Closed string channel} \\ \hline 
Coupling & Usual gauge coupling & v.e.v. of bulk field $\phi $ at $x_{\perp 
} $ \\ \hline 
Running & Propagation of SM nonsinglet states & $\phi $ propagator in bulk 
\\ \hline 
Logarithmic; $\alpha ^{-1}\sim \ln (\Lambda R)$ & D=4 propagation & 
$d_{\perp }=2$ propagation \\ \hline 
Linear: $\alpha ^{-1}\sim (\Lambda R)$ & D=5 propagation. Single KK tower. &  
$d_{\perp }=1$ propagation \\ \hline 
Quadratic $\alpha ^{-1}\sim (\Lambda R)^{2}$ & D=6 propagation. Double 
``in-brane'' KK tower & $d_{\perp }=0$ \\ \hline 
String cut-off & UV, $M_{s}$ or $M_{winding}\geq M_{s}$ & IR, Lightest $\phi  
$. \\ \hline 
Coupling constant ratios & Gauge symmetry & Geometric symmetry  in the
brane sector
\\ \hline 
\end{tabular} 
\caption{Alternative description of gauge coupling running in open/closed 
string channels.\label{table:1}}%
\end{table}%
%
 
\subsection{\it Power-law unification} 
 
The first discussion \cite{dienes_v1} of the possibility of a low string and 
unification scale following from power law running employed an effective 
field theory approach. In this the individual logarithmic corrections of 
Kaluza Klein modes (with respect to $\delta=(p-3)$ ($p\geq 3$) ``in-brane'' 
dimensions) below the string scale are summed \cite{taylor, 
string_thresholds}. This leads to the following (one-loop) RGE evolution  
\cite{dienes_v1}  
\begin{eqnarray} 
\alpha _{i}^{-1}(\mu ) &=&\alpha ^{-1}(\Lambda
)+\frac{\overline{b}_{i}}
{2\pi }\sum_{\vec{m}}^{|\vec{m}|\leq \Lambda /\mu _{0}}\ln
\frac{\Lambda }
{\mu _{0}|{\vec{m}}|}+\frac{b_{i}}{2\pi }\ln \frac{\Lambda }{\mu } 
\label{low_mstring} \\ 
&\approx &\alpha ^{-1}(\Lambda )+\frac{{\overline{b}_{i}}}{2\pi }\left\{
\frac{
\pi ^{\delta /2}}{\delta \;\Gamma (1+\delta /2)}\left[ \left( \frac{\Lambda  
}{\mu _{0}}\right) ^{\delta }-1\right] -\ln \frac{\Lambda }{\mu _{0}}
\right\} +\frac{b_{i}}{2\pi }\ln \frac{\Lambda }{\mu } \label{low2_mstring}
\end{eqnarray} 
which shows explicitly \cite{string_thresholds} 
the origin of the ``power-law'' behaviour in the high 
degeneracy of Kaluza-Klein modes. Here ${\overline{b}_{i}}$ is the $N=2$ 
beta function of KK sector and $b_{i}$ is the one loop contribution, $\delta  
$ is the effective number of (``in-brane'') extra dimensions, 
$\Lambda $\ is the cut-off 
scale and $\mu _{0}$ is the mass of the lowest KK excitation. Although the 
number of extra dimensions may be as large as six, as discussed above, we 
expect at most quadratic running, $\delta =1,2$. The constraint (\ref{weak}) 
restricts the in-brane volume to values close to unity in string length 
units. In general the power-law running takes place over a very small range 
of energy and thus the radii ($1/\mu _{0}$) are close to the string scale 
and (\ref{weak}) may be satisfied by symmetric unifications with the 
remaining $6-\delta $ dimensions also of order the string length scale. In 
the effective field theory approach unification occurs at the cutoff
scale 
$\Lambda $ and this scale must be identified. The usual assumption is to 
identify it with the string scale, $M_{I}.$ However, following the 
discussion above, in Type I$^{\prime }$ theories it may be one should 
identify $\Lambda $ with the larger winding mode mass. The correct 
identification depends on the details of the string theory - in some cases 
the winding mode excitations fill out complete $N=4$ multiplets and in this 
case the lower string scale is the appropriate choice \cite{dudash}. In 
either case the power law running can lead to unification at a scale which 
may be smaller \cite{dienes_v1} than the ordinary MSSM unification scale. 
 
While power law running can reduce the unification scale there are two 
difficulties in obtaining the usual MSSM predictions. The first is the 
sensitivity power law running introduces to the thresholds at the 
unification scale \cite{extradim}. This is readily seen from 
eqs.(\ref{low_mstring}), (\ref{low2_mstring})
where, for a low (TeV) unification scale $\mu _{0}(\partial 
\alpha _{i}^{-1}/\partial \mu _{0})\simeq \delta (\alpha _{i}^{-1}(\mu 
)-\alpha ^{-1}(\Lambda )).$ This is to be compared to the case of normal 
logarithmic running in the MSSM where $\mu _{0}(\partial \alpha 
_{i}^{-1}/\partial \mu _{0})\simeq (\alpha _{i}^{-1}(\mu )-\alpha 
^{-1}(\Lambda ))/\ln (\Lambda /\mu ).$ Thus the sensitivity to the 
unification scale is roughly enhanced by a factor 
$\delta \ln (\Lambda /\mu )$ which 
is $\geq 32.$ A more careful analysis gives $\delta\mu_o/\mu_o=1/224$
\cite{extradim}.
This means that in order to make a precision prediction for 
the gauge coupling one must know the detailed spectrum of the states at the 
unification scale. Although this does not rule out the possibility of low 
scale unification through power law running, it does make the success of the 
detailed prediction of the gauge couplings in the MSSM seem surprising. 
 
The second difficulty one encounters in trying to duplicate the MSSM results 
via power law running comes from the fact that in eq.(\ref{low_mstring}) the 
rate of running is controlled by the coefficients ${\overline{b}_{i}}$ which 
are determined by the $N=2$ sector of the theory. The difficulty is that 
these are not expected to be proportional to the $N=1$ beta functions and 
thus we may expect substantial deviations from the MSSM predictions. To see 
this note that if one has towers of KK states for the entire MSSM
spectrum,
${\overline{b}_{i}}=-2T_{i}(G)+2\sum_{\psi }^{{}}T_{i}(\psi )$ which is not 
proportional to $b_{i}^{MSSM}=-3T_{i}(G)+\sum_{\psi }^{{}}T_{i}(\psi )$ of 
the ($N=1$) MSSM sector. Instead we have ${\overline{b}_{i}}
=2b_{i}^{MSSM}+4T_{i}(G)$. To avoid this we should look for a model for 
which  
\begin{equation} 
{\overline{b}_{i}}=Kb_{i}^{MSSM}+C  \label{kaku} 
\end{equation} 
with $K$ and $C$ some gauge group independent constants. In this case 
eq.(\ref {low_mstring}) gives the same value 
for $\alpha _{3}(M_{z})$ as in the MSSM 
at one loop level (in such a case we would have an explanation for the 
``accidental'' success of the MSSM gauge unification predictions). 
 
Attempts to obtain a spectrum satisfying this condition have been made by 
Kakushadze \cite{zurab1} by including further KK states which, together with 
the KK excitations of the MSSM, have one loop coefficients which
satisfy eq.(\ref{kaku}). 
A simple example with $K=1$\cite{zurab1} was given by a $Z_{2}$ 
orbifold model which breaks an underlying ${N}=2$ supersymmetry to ${N}=1$. 
The ${N}=2$ spectrum is that of the MSSM KK excitations together with those 
of additional superfields $F_{\pm }$, $SU(3)\times SU(2)$ singlets
with $U(1)_{Y}$ charge $\pm 2$ 
respectively.  However the addition of the $F_{\pm },$ while leading 
to a the $N=2$ beta function which satisfies eq.(\ref{kaku}), 
does not reproduce the $N=1$ MSSM spectrum because it contains additional  
{\it light} ${N}=1$ fields with the quantum numbers of $F_{\pm }$. As a 
result one finds  
\begin{equation} 
\alpha _{3}^{-1}(M_{z})=\alpha _{3}^{o-1}(M_{z})+\frac{1}{2\pi }\frac{6}{7} 
\ln \frac{M_{s}}{M_{F}}  \label{Z1} 
\end{equation} 
Here $\alpha _{3}^{0}(M_{z})$ is the MSSM (one loop) value while the last 
term is the (one-loop) contribution of $F_{\pm }$ states to the running of 
the gauge couplings. This term depends on the (unknown) ratio $M_{s}/M_{F}$ 
and thus eq.(\ref{Z1}) makes no prediction for $\alpha _{3}(M_{z})$ 
following from the unification condition. A detailed mechanism or further 
input is needed to fix the value $M_{s}/M_{F}$. Thus, although the model may 
allow the presence of a low (TeV) string scale, it does not {\it
predict} $\alpha _{3}(M_{z})$. 
It may be that one may construct models of this type 
which do not modify the $N=1$ spectrum but one may see from this example 
the difficulty in accommodating the MSSM predictions in the 
framework of power-law running. 
 
\subsection{\it Logarithmic unification} 
 
We turn now to the possibility that the gauge couplings unify with 
logarithmic running. A Type I/I$^{\prime }$ full string calculation for the 
threshold corrections to the gauge couplings was performed for the case of 
an $N=2$ orientifold based on the $T^{4}/Z_{2}$ orbifold 
\cite{fabre,partouche} and other $N=2$ orientifolds obtained by toroidal 
compactification of six-dimensional vacua \cite{dudash}. These results were 
also generalised \cite{dudash} to four dimensional compactifications
with 
$N=1$ supersymmetry. In the latter case the cut-off for $N=2$ sector is not 
the string scale itself but the first winding mode above the string scale 
associated with the (``off-the-brane'') two-torus while for the light states  
$N=1$ the cut-off is set by the string scale itself. Two possibilities exist 
for the RGE behaviour \cite{dudash,bachas} at the high scale in this case. 
One of these is the linear ``power-law'' regime, corresponding to an 
``asymmetric'' two-torus compactification, while the other has only 
logarithmic dependence on $R$ (and ultimately on $M_{P}$) in the RGE. We 
will now concentrate on the second case as the first one was discussed above. 
 
The interest for phenomenology in these models is two-fold. Firstly, the 
large transverse volume and low string scale brings the possibility (e.g. 
for the case the MSSM is confined to a D3 brane) of new gravitons and their 
Kaluza Klein towers in the $mm^{-1}$ range \cite{savas}, a situation not 
ruled out by the experiment. Secondly, the mild logarithmic dependence of 
the couplings on the high scale raises the possibility \cite{bachas} of 
gauge coupling unification at a large scale above $M_{I}$ given by the 
string cut-off (first winding mode). This may be just the original 
unification scale of the MSSM thus elegantly providing an alternative (Type 
I) interpretation of the successful MSSM predictions. 
 
Turning to a more detailed discussion of the latter possibility note that 
the corrections to gauge couplings come only from the ${N}=1$ (massless) 
sector and from the ${N}=2$ massive winding (Type I$^{\prime }$) (or Kaluza 
Klein, Type I) sector. Let us consider the ${N}=2$ sector first. The massive  
${N}=2$ threshold corrections to $\alpha ^{-1}$ at the string scale $M_{I}$ 
have the form \cite{dudash} (for both ${\cal A}$ (annulus) and ${\cal M}$ 
(M\"{o}bius strip) geometries relevant at one loop)  
\begin{equation} 
\Delta _{i}^{TypeI}=-\frac{1}{4\pi }\sum_{m}{\overline{b}_{i,m}}\left\{ \ln 
\left( \sqrt{G_{m}}ImU_{m}M_{I}^{2}|\eta (U_{m})|^{4}\right) \right\} 
\label{typeI} 
\end{equation} 
Here $G_{m}$ is the metric on the torus\footnote{
We consider in the following only compactification on a single torus, 
$T^{1}$.} $T^{m}$, and $\sqrt{G_{1}}=\tilde{R}_{1}\tilde{R}_{2}$ (for 
a rectangular torus). The behaviour of the thresholds\footnote{For 
Type I$^{\prime }$ case we should take $R=1/{\tilde{R}}$} when 
$ImU={\tilde{R}_{1}}/{\tilde{R}_{2}}$ is of order one is indeed 
logarithmic, $\Delta _{a}\sim \ln ({\tilde{R}_{1}}{\tilde{R}_{2}})$. 
 
We note that unlike the weakly coupled heterotic case \cite{dixon} where 
quadratic dependence on the common radius of the two torus (associated with 
the $N=2$ sector) exists, in the present case this does not happen. The 
condition of global tadpole cancellation ensures that after adding the 
contributions (each regulated by an infrared cut-off) of massless closed 
string states emitted into the transverse channel, the two $R^{2}$ dependent 
results obtained for ${\cal A}$ and ${\cal G}$ cancel. Therefore there is no 
quadratic term present in (\ref{typeI}). In this calculation the ultraviolet 
(would-be quadratic) behaviour in the open channel is regulated by the 
infra-red regime of the closed string channel (c.f. Table~1). 
 
From eq.(\ref{typeI}), after adding the usual ${N}=1$ one loop term 
proportional to $b_{i}$ we have  
\begin{equation} 
\alpha _{i}^{-1}(M_{z})=\alpha _{I}^{-1}+\frac{b_{i}}{2\pi }\ln 
\frac{M_{I}}{M_{z}}+\frac{\overline{b}_{i}}{2\pi }\ln 
\frac{\tilde{\mu}}{M_{I}} 
\label{logrunning} 
\end{equation} 
where we have taken ${\tilde{R}_{1}}={\tilde{R}_{2}}$ 
and $\tilde{\mu}$ =$1/({\tilde{R}_{1}}|\eta (U_{m})|^{2}).$ 
Here ${\overline{b}_{i}}$ is the ${N}=2$ 
one-loop beta function coefficient. 
 
From this equation we can immediately identify a problem in 
obtaining the usual MSSM results with a low string scale. As in the case of 
power law running, the dominant evolution is governed by the $N=2$ sector. 
We therefore have a similar difficulty to that discussed above in obtaining 
the MSSM results because in general ${\overline{b}_{i}}$ is not proportional 
to $b_{i}^{MSSM}.$ In ref.\cite{jmr} an orbifold projection was used to 
break $N=2$ supersymmetry to $N=1$. In this case the running up to the 
unification scale (i.e. first winding mode scale $\tilde \mu$) 
is indeed given by the 
$N=1 $ sector apparently solving this problem. However the 
construction works only for modding by freely acting groups and this 
implies the $N=1$ matter fields giving rise to ${\overline{b}_{i}}$ 
originate from the adjoint representation of the underlying gauge 
group before modding. 
Although in the orbifolded model there are $N=1$ matter fields 
transforming as the fundamental representations $F$ under one of 
the group factors, there are also fields transforming as ${\overline
F}$ \cite{trivedi}. This does not allow for an identification of 
these fields with the MSSM matter fields.
 As a result  we still have the problem of explaining why 
the ${\overline{b}_{i}}$ should 
satisfy eq.(\ref{kaku}). If we assume however that this problem can be 
solved, eq.(\ref{logrunning}) together with eq.(\ref{kaku}) reproduces the 
MSSM unification at one loop level with $(\tilde{\mu}/M_{I})^{K}M_{I}\approx 
10^{16}$ GeV. For the case $K=1$ the unification occurs at $\tilde{\mu}
\approx 10^{16}$ GeV. 
 
Somewhat surprisingly logarithmic unification also has a fine tuning 
problem. Naively our previous discussion suggests logarithmic unification 
will be no more sensitive to threshold effects than the MSSM. However this 
is misleading because here the string cut-off scale $\tilde{\mu}\approx 
10^{16}\,GeV$ corresponds to a cross channel exchange state with mass 
$M_{I}^{2}/\tilde{\mu}$. For a 1 TeV string scale this is $10^{-10}$ $GeV$ 
and corresponds to the lightest mass of Kaluza Klein modes in the closed 
string channel. However the state controlling the UV behaviour of the open 
string channel has vacuum quantum numbers. Such a state has no symmetry 
(local gauge symmetry etc) to protect it from receiving contributions to its 
mass at scales below the supersymmetry breaking scale. We therefore expect 
that such a state will acquire mass from supersymmetry breaking effects. The 
magnitude of such a mass depends on the supersymmetry breaking mechanism but 
has a lower bound governed by the supersymmetry breaking communicated by 
gravitational effects. If supersymmetry is broken in a hidden sector (e.g. 
on another brane) and communicated to the visible sector by\ gravitational 
effects then we know such effects must be of $TeV$ scale otherwise some of 
the SUSY\ partners of the SM should have been seen. However the 
supersymmetry breaking in the bulk must be at least of $TeV$ scale as well 
because, in this case, SUSY\ breaking is communicated to the visible sector 
via the bulk. As a result logarithmic running is {\it not} a viable 
mechanism for gauge unification  because
now the closed string state mass $(M_I^2/\tilde{\mu})\approx 1 TeV$ 
so with $M_I\approx 1 TeV$ we have  $\tilde{\mu}\simeq 1TeV$
instead of the  $10^{16}$ GeV needed for unification. 
If, to avoid this problem, one requires the 
closed string exchange state with mass $M_{I}^{2}/\tilde{\mu}$ be
$\geq $ 
$TeV$ scale, while maintaining high scale unification ($\tilde{\mu}\approx 
10^{16}\,GeV)$ one finds the string scale is bounded by $M_{I}\geq 10^{10}$ 
GeV. 
 
The only possibility to maintain logarithmic unification with a low string 
scale is that supersymmetry breaking in the visible sector is much larger 
than in the bulk. This happens in gauge mediated scenarios where the 
supersymmetry breaking is communicated to the visible sector by gauge 
interactions much stronger than gravity. Such a scheme requires that 
supersymmetry breaking occurs on the SM brane itself. However, even in such 
a scheme there is a problem in keeping the string scale as low as $1TeV$ as 
gravitational effects still generate a supersymmetry breaking mass of order 
the gravitino mass for fields in the bulk. The gravitino mass is of
order $M_{Susy}^{2}/M_{P}$ where\footnote{We denote by $M_P$
the Planck scale.} $M_{Susy}$ is the supersymmetry  breaking scale on 
our brane. Demanding that the gravitino mass  
is less than $10^{-10}$ $GeV$ to avoid the fine 
tuning problem implies $M_{Susy}\leq 10^{4}GeV.$ Provided one remains in the 
perturbative domain this is probably too small for a viable gauge mediated 
scheme because SUSY breaking in the visible sector is given by 
$\gamma M_{Susy}^2/M$ where $\gamma $ is the effective coupling 
of the visible sector to 
the SUSY breaking sector and $M$ is the messenger mass 
$(M\geq M_{Susy})$. 
In viable models it occurs at one loop order
for gauginos \cite{giudice} suggesting $\gamma $ is too small to generate 
acceptable masses for the MSSM sparticle spectrum. 
Even if it proves to be possible one sees 
that the resulting model is heavily constrained requiring gauge mediated 
supersymmetry breaking with the gauge mediator mass, $M$, in the $TeV$
range too. Gauge mediation  with a light messenger sector 
$(< 10^5 GeV)$ is however unnatural \cite{strumia}. 

To summarise, in both the case of power law and logarithmic running there 
are difficulties in accommodating the success of the MSSM unification 
predictions in models with a low string scale. In the case of power law 
unification there is also an enhanced sensitivity to the details of the mass 
spectrum at the unification scale making the success of the detailed 
prediction of the gauge couplings in the MSSM quite surprising. In the case 
of logarithmic running there is a fine tuning problem which can 
be alleviated 
by raising the string scale significantly above the $TeV$ scale. However in 
both cases it is difficult to generate the MSSM beta function via the N=2 
sector. To avoid this difficulty in these models it is thus 
necessary to give up the MSSM structure. From this point of view,
in this case the MSSM remarkable success in predicting gauge 
unification appears to be accidental. To illustrate this last 
possibility we consider unification in a promising Type I which has a 
multiplet structure close to that of the MSSM. 
 
\subsection{\it A Type I model with intermediate unification} 
 
The model we wish to consider is a specific example of a larger class
of $D=4 $ Type I vacua obtained from a standard $D=4,$ $N=1$ compact 
Type IIB 
orientifold with D$_{p}$ branes and anti-D$_{p}$ branes located at different 
points of the underlying orbifold \cite{quevedo}. We consider here the class 
of models of \cite{quevedo} which have gravity mediated supersymmetry 
breaking with the full Standard Model embedded in a 9-brane sector and with 
an additional set of 5-branes trapped at the fixed points of the $Z_{3}$ 
orientifold. This model has no $N=2$ sector and the UV cutoff scale
for the $N=1$ sector is the string scale \cite{dudash}. As we shall 
discuss this model 
can give gauge unification via the radiative corrections of the $N=1$ sector 
with an intermediate string scale and without introducing significant 
threshold sensitivity. The price to achieve this is to depart from the MSSM 
light matter content and to start with a non-MSSM relation between the gauge 
couplings. In particular the normalisation of the hypercharge of these 
states is $\alpha _{Y}/\alpha _{s}=3/11$ at the string scale (instead of the 
usual ``SU(5) factor'' $3/5$). The massless spectrum contains the spectrum 
of the MSSM. In addition there are five further pairs of Higgs doublets and 
three vector-like right handed colour triplets ($d^{c}+{\overline{d^{c}}}$). 
The remaining states of the model may have mass of the order of string 
scale. In determining the unification prediction of this model we follow  
\cite{quevedo} and assume that these additional states have a common 
{\it bare} mass\footnote{ 
For the purpose of two loop RGE only the bare mass of the fields is 
needed.}, $\tilde{m}$, which must not be greatly different from the 
electroweak scale. 
 
Including the threshold effects at $\tilde{m}$, the two loop RGE equations 
are given by  
\begin{equation} 
\alpha _{i}^{-1}(M_{z})=-\delta _{i}+\alpha
_{s}^{-1}+\frac{b_{i}}{2\pi }\ln  
\left[ \frac{M_{s}}{M_{z}}\right] +\frac{\delta b_{i}}{2\pi }\ln \left[  
\frac{M_{s}}{\tilde{m}}\right] +\frac{1}{4\pi }\sum_{j=1}^{3}Y_{ij}\ln \left[ 
\frac{\alpha _{g}}{\alpha _{j}(\tilde{m})}\right] +\frac{1}{4\pi }
\sum_{j=1}^{3}\frac{b_{ij}}{b_{j}}\ln \left[ \frac{\alpha _{s}}{\alpha 
_{j}(M_{z})}\right]  \label{HMSSM} 
\end{equation} 
where  
\begin{equation} 
Y_{ij}=\frac{\delta b_{j}}{b_{j}b_{j}^{\prime }}\left[ 2b_{j}T_{j}(G)\delta 
_{ij}-b_{ij}\right]  \label{why} 
\end{equation} 
with $T_{j}(G)=\{0,2,3\}_{j}$, $b_{ij}$ is the two loop beta function as in 
the MSSM but with $3/11$ hypercharge normalisation, $b_{j}=\{11\xi 
,1,-3\}_{j}$, ($\xi =3/11$), $b_{j}^{\prime }=b_{j}+\delta b_{j}$,
with 
$\delta b_{j}=\{7\xi ,5,3\}$ to account for the additional five Higgs pairs 
and three pairs of right handed colour triplets. For the terms $Y_{i3}$ 
equations (\ref{HMSSM}) should be understood as {\it limits} of $%
b_{3}^{\prime }\rightarrow 0$ which indeed give the appropriate (finite) 
results when multiplied by the term $\log (\alpha _{g}/\alpha
_{3}({\tilde{m}
}))$. Finally one should include the low energy supersymmetric thresholds $%
\delta _{i}$ and regularisation scheme conversion factors (${\overline{MS}}%
\rightarrow {\overline{DR}}$). Usually $\delta _{i}$'s are included as an 
overall effect on $\alpha _{3}(M_{z})$ through an {\it effective} 
(low-energy) supersymmetric threshold \cite{langacker}. In this case the 
different hypercharge normalisation of $\delta _{1}$ prevents us from using 
this approach, although their overall effect on $\alpha _{3}(M_{z})$ is not 
expected to change significantly from the ``normal'' hypercharge case 
(when $\alpha _{3}(M_{z})$ is reduced by an amount of $\approx 2-4\%$). 
 
The predictions for the string coupling  and the 
string scale $M_{s}$ can be read from the plots in Figure 1 to see 
that the model indeed allows for a logarithmic unification at about $\approx 
10^{10}M_{z}$ GeV for a strong coupling consistent with the experimental 
value \cite{particledata} $\alpha _{3}(M_{z})=0.119\pm 0.002$. This requires 
an intermediate threshold at about\footnote{The results are 
subject to additional thresholds at the string scale, which may change this 
picture significantly.} ${\tilde m} \approx (100-300)M_{z}$.

One can also compute  the bottom to tau mass ratio, taking account of the 
different normalisation of $\alpha_Y/\alpha_s$. For the model with
the above spectrum set-up, we find  the following one loop level
result (including only gauge effects)
\begin{equation}
\frac{R(M_s)}{R(M_z)}=
\left(1+\frac{\alpha_s b_1'}{2\pi}\ln \frac{M_s}{\mu_o} 
\right)^{\frac{1}{b_1'}\frac{2 N_1 \delta b_1}{b_1} }
\left(\frac{M_s}{\mu_o}\right)^{ {\frac{\alpha_s}{2\pi}}\frac{2 N_3
\delta b_3}{b_3} }
\left(\frac{\alpha_s}{\alpha_1(M_z)}\right)^{-\frac{2 N_1}{b_1}}
\left(\frac{\alpha_s}{\alpha_3(M_z)}\right)^{-\frac{2 N_3}{b_3}}
\end{equation}
where $N_1=-10/9 \xi$, $N_3=8/3$ and $R(Q)$ denotes 
the ratio bottom to tau mass at the scale Q. 
For $\alpha_s\approx 0.1$ (needed to generate $\alpha_3(M_z)$) we find that 
$R(M_z)\approx 5 R(M_s)$. If we assume bottom-tau unification 
at the string scale, this implies $m_b(M_z)\approx 8.731GeV$. 
To correct this, model dependent Yukawa effects not considered 
here must be very significant, but then the MSSM result for 
the bottom-tau mass ratio should  be regarded as yet another
accident.
 
This example shows that it is possible in a realistic model to obtain good 
results for gauge unification in a manner incompatible with the original 
MSSM, demonstrating that the latter's success may indeed be an illusion. 
However it also illustrates the difficulties a non-MSSM scheme must surmount 
in order to approach the accuracy of the MSSM prediction. Although the 
unification is logarithmic the sensitivity to thresholds is considerably 
enhanced due to the light states additional to the MSSM. For example the 
uncertainty in the additional Higgs doublets mass translates to the 
uncertainty $\delta \alpha _{2}^{-1}=(5/2\pi )\delta M_{H}/M_{H}.$ If one is 
to predict $\alpha _{3}(M_{z})$ to the present experimental accuracy then 
one must know these masses to better than $\delta M_{H}/M_{H}=0.1.$ In
the  light of this the precision of the MSSM prediction for the 
strong coupling  seems even more 
remarkable and makes it harder for us to accept it is just an accident.

\section{Conclusions} 
 
We have considered the possibility that Type I string theories with a low 
(TeV) string scale may accommodate the successful MSSM predictions for the 
unification of the gauge couplings. Our conclusions are pessimistic. Models 
with power-law running have a significant sensitivity to the unification 
scale thresholds. Models with logarithmic running up to the string cut-off 
(winding) scale far above the string scale reduce the sensitivity to 
this scale but have a fine tuning problem associated with the closed string 
spectrum which requires either a very special Supersymmetry breaking sector 
or an intermediate string scale. In both cases it is difficult to generate 
the MSSM beta function via the N=2 sector. Of course it may be that the 
success of the MSSM predictions is illusory and the existence of a realistic 
Type I theory which makes a non-MSSM prediction for the gauge couplings 
consistent with the measured values demonstrates this possibility. However, 
due to the threshold dependence on the non-MSSM states, it is not possible 
in this model to make a prediction for the gauge couplings to the same 
accuracy as the MSSM result without a detailed knowledge of the masses
at the unification scale.
All this emphasizes the fact that if the MSSM result is regarded as 
an accident it is a surprisingly accurate accident! In the light of 
all this it seems to us that  the original heterotic string 
(possibly strongly coupled), being able 
naturally  to accommodate the MSSM results, still offers the 
best explanation of the unification of gauge couplings. 
 
\section{Acknowledgments} 
We would like to thank I. Antoniadis and L. Ib\'{a}\~{n}ez
for useful discussions. 
D.G. acknowledges the financial support from the part of the University of 
Oxford (Leverhulme Trust research grant). 
The research is supported in part by the 
EEC under TMR contract ERBFMRX-CT96-0090. 

\begin{figure}
\begin{tabular}{cc|cr|}
\parbox{8cm}{\psfig{figure=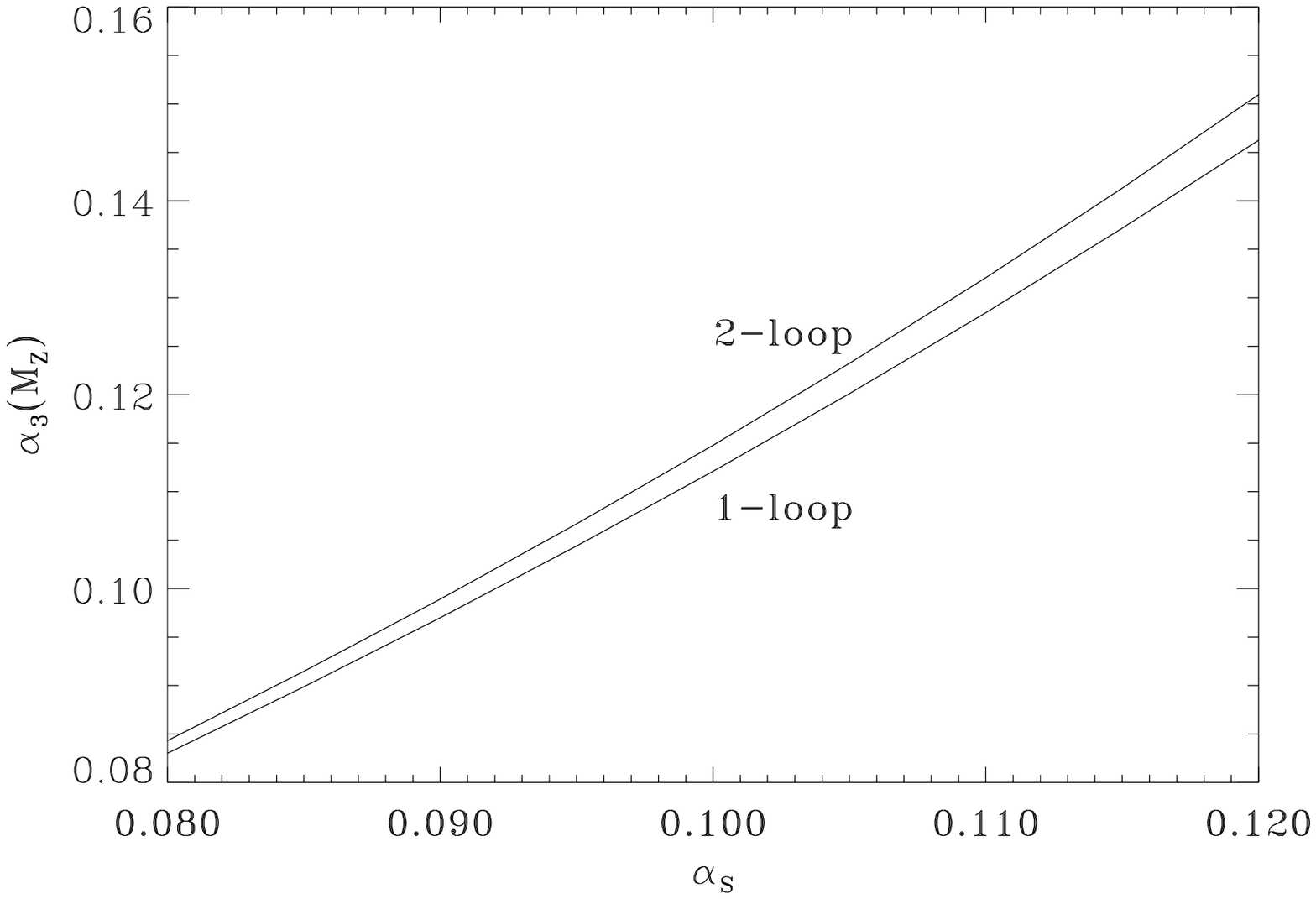,height=7.5cm,width=7.5cm}
}\hfill{\,\,\,\,\,\,\,\,\,\,\,\,\,\,\,}\
\parbox{8cm}{\psfig{figure=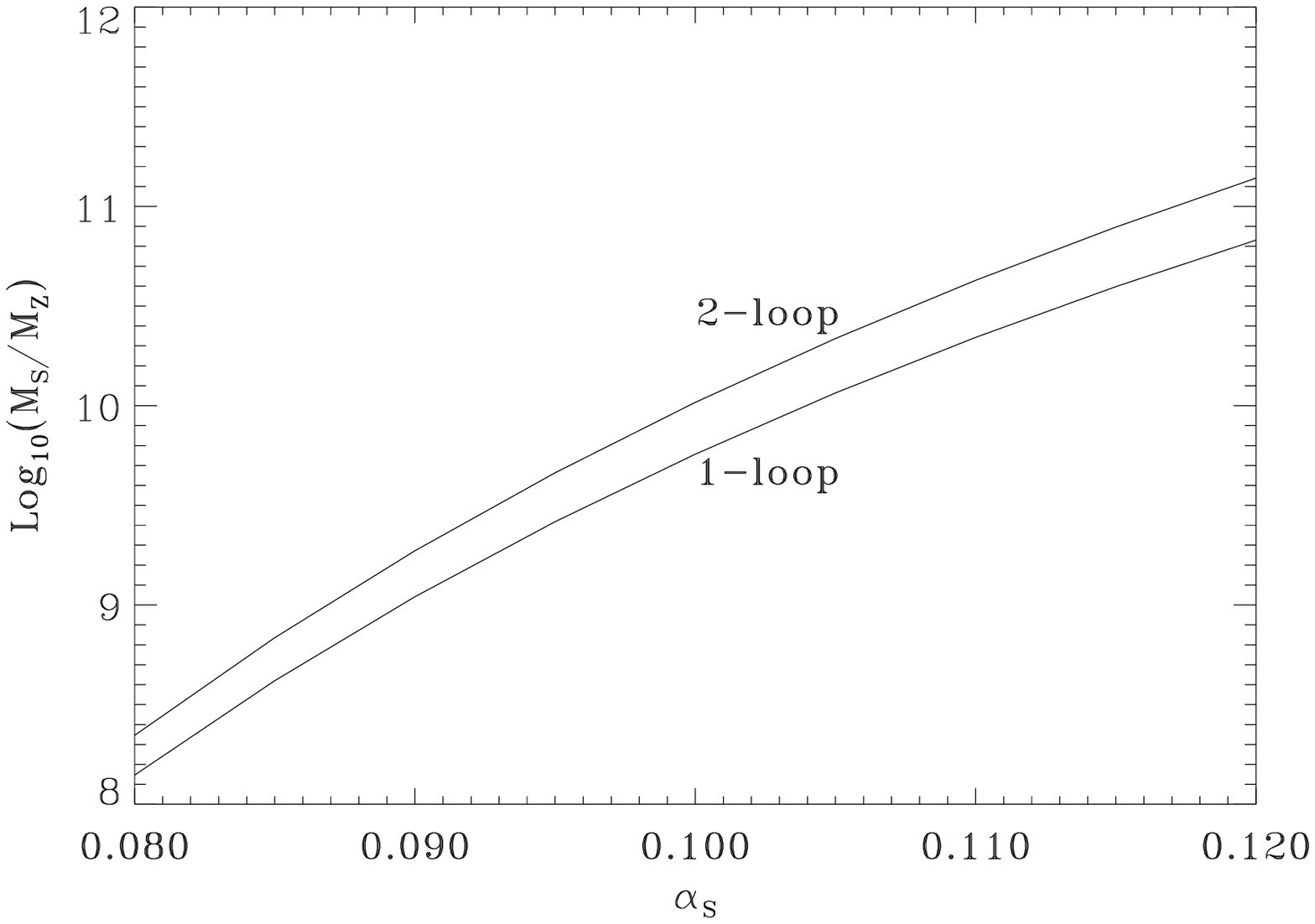,height=7.5cm,width=7.5cm}}
\end{tabular}
\caption{The values of $\protect\alpha_3(M_z)$ (left)
and $\log_{10}(M_s/M_z)$ (right) as a function of 
the string coupling $\protect\alpha_s$. In both cases the lower curve  
is a one loop approximation while the upper curve  is the two loop 
case. For $\protect\alpha_3(M_z)\approx 0.12$ we require 
$\protect\alpha_s\approx 0.1$. The overall picture 
for $\protect\alpha_3(M_z)$ is expected 
to be shifted downwards by about $2-4 \% $ after the inclusion of 
the low energy threshold corrections $\protect\delta_i$
with the appropriate hypercharge normalisation, 
while the  remaining plot is not expected to change significantly.}
\end{figure}

\end{document}